\documentstyle{article}
\begin{document}
\newcommand{\newc}{\newcommand}

\newc{\be}{\begin{equation}}
\newc{\ee}{\end{equation}}
\newc{\ba}{\begin{eqnarray}}
\newc{\ea}{\end{eqnarray}}
\newc{\bann}{\begin{eqnarray*}}
\newc{\eann}{\end{eqnarray*}}
\newc{\ie}{{\it i.e.}}
\newc{\eg}{{\it eg.}}
\newc{\etc}{{\it etc.}}
\newc{\etal}{{\it et al.}}

\newc{\ra}{\rightarrow}
\newc{\lra}{\leftrightarrow}
\newc{\lsim}{\buildrel{<}\over{\sim}}
\newc{\gsim}{\buildrel{>}\over{\sim}}

\begin{titlepage}
\begin{center}
January 1995\hfill
CfA Preprint No. 4019 \\
\hfill  astro-ph/9502004 \\
             \hfill
\vskip 1in

{\large \bf
Spectra and Statistics of Cosmic String Perturbations on the
Microwave Background: A Monte Carlo Approach
}

\vskip .6in
{\large Andrea M. Gilbert}\footnote{E-mail address:
agilbert@cfa.harvard.edu} and {\large Leandros
Perivolaropoulos}\footnote{E-mail address:
lperivol@cc.uoi.gr}
\vskip 0.5cm
{\em Harvard-Smithsonian Center for Astrophysics\\
 60 Garden St.\\
Cambridge, MA  02138, USA.}\\[.15in]
\end{center}
\vskip .7in
\begin{abstract}
\noindent
	Using Monte Carlo simulations of perturbations induced by
cosmic strings on the microwave background, we demonstrate the scale
invariance of string fluctuation patterns.  By comparing
string-induced fluctuation patterns with gaussian random phase ones,
we show that the non-gaussian signatures of the string patterns are
detectable by tests based on the moments of the distributions only for
angular scales smaller than a few arcminutes and for maps based on the
gradient of temperature fluctuations.  However, we find that tests of
the gaussianity of the moments fail when we include a reasonable
amount of instrumental noise in a pattern.  Signal to noise ratios of
$3.3$ or greater completely suppress a string pattern's non-gaussian
features even at the highest resolutions.
\end{abstract} \end{titlepage}

\section{Introduction}
	The major progress made during the last decade in both
cosmological observation and theory has made the search for
understanding of the origin of cosmic structure one of the most
exciting fields of scientific research.

	Two classes of theoretical models explaining the origin of
structure have emerged during the last decade and have survived with
only minor adjustments through strong constraints imposed by detailed
observational data.  According to the first class of models, the
primordial fluctuations that gave rise to structure in the universe
were produced by quantum fluctuations of a scalar field during a
period of exponential growth of the size of the universe.  This period
is known as {\it inflation} and was invented in order to causally
explain the observed large scale isotropy and homogeneity of the
universe.  The primordial fluctuations produced during inflation may
be shown to obey Gaussian statistics and to have a {\it scale
invariant} spectrum.  In this context scale invariance means that the
amplitude of fluctuations at the time when they enter the causally
connected horizon is fixed ({\it i.e.} independent of cosmic time).

	According to the second class of theories, primordial
fluctuations are provided by {\it seeds} of trapped energy density produced
during a phase transition in the early universe.  Such plase
transitions are predicted by particle physics theories known as Grand
Unified Theories (GUT's), which attempt to unify all known
interactions under a single theory.  The transition is predicted to
occur at $10^{-35}$ seconds after the Big Bang, when the temperature
of the universe is about $10^{16} GeV$.

	The seeds of energy produced during the phase transition are
called {\it topological defects} and are effectively localized regions
where the symmetric high temperature phase of the universe has been
trapped and its decay to the present low temperature vacuum manifold
is prevented by topological considerations.  According to their
geometry, topological defects are classified into monopoles (stable
pointlike defects), cosmic strings (stable linear defects), domain
walls (planar defects), and textures (collapsing pointlike defects).
Most theories for large scale structure formation based on monopoles
and domain walls face severe conflicts with observations and are not
favored \cite{b94}.  On the other hand, theories based on cosmic
strings and textures show several encouraging features when compared
with observations and are among the main contendors proposed to
explain the origin of structure in the universe.

	In what follows, we focus on observational tests for theories
based on cosmic strings.

	The formation and evolution of strings is described in a
review by Vilenkin \cite{av85}.  It has been shown, using both
analytical arguments \cite{av85} and numerical simulations
\cite{simul} of string network evolution, that the network of strings
approaches a configuration which looks the same at all times with
respect to the causal horizon scale.  This configuration is known as
the {\it scaling solution}.  According to this solution the string
network at any given time consists of about $10$ long strings with
radii of curvature of the order of the causal horizon scale, along
with a component of small string loops with radii much smaller than
the horizon scale
\cite{simul}.

	The presence of such a network in the universe results in
several observational predictions which can be used to test the
underlying theory.  The most important prediction is that of large
scale structure formation.  At early times, the cosmic string network
produces density and velocity perturbations in the surrounding matter.
These perturbations grow gravitationally and can lead to the presently
observed structure in the universe.  It has been shown \cite{tb86,pbs90}
that the mass per unit length ($\mu$) of strings required in order to get
bound structures by today is
\be G\mu \simeq {10^{-6}} \ee
where $G$ is Newton's constant, included to obtain a dimensionless
number.  This condition is consistent with requirements imposed by
microphysics in order to make GUT's consistent with particle physics
accelerator experiments \cite{pl81,dp84}.  If this condition is
satisfied, the cosmic string model predicts the existence of the
following: filaments and sheets of galaxies (formed in the wakes of
moving long strings) with thicknesses comparable with observations
\cite{pbs90,v86,s87,vv}; coherent large-scale velocity fields
 \cite{v92a,pv94}; perturbations on the microwave background sky
\cite{gs}; and galactic magnetic fields \cite{v92b}.  It also predicts
the existence of ultra-high energy gamma ray events \cite{mb93,bst93},
gravitational lensing of quasars \cite{vu91}, and a stochastic gravitational
radiation background \cite{ca92}.  Despite the large number of
predictions made by the string model and the detailed existing data,
it has not been possible to rule out the model using any particular
observation.  This is due in part to the complexity of the model,
which introduces large uncertainties in its predictions, but it may
also be due to features of the model which could contain elements of
the physically realized theory.

	The recent detection of fluctuations on the $2.7 K$ microwave
background sky, first by the COBE collaboration on angular scales of
$10^\circ$ \cite{gs} and later by other experiments on smaller angular
scales, has provided a powerful new tool for testing theoretical
models.  The main source of string-induced fluctuations on the
microwave background at large angular scales is the {\it
Kaiser-Stebbins effect}.

  	According to the Kaiser-Stebbins effect, a moving long string
present between the last scattering surface (the horizon at the time of
recombination $t_{rec}$) and the present time $t_0$ produces
temperature discontinuities between photons reaching the observer from
opposite sides of the string (Figure $1$).  These discontinuities have
the form of a step function \cite{ks,s}, and are caused by the
geometry of spacetime around the string; the space around a straight
long string is locally flat, but globally it has the geometry of a
cone with a deficit angle equal to $8\pi G \mu$ \cite{vdef}.  Thus, a
moving long string interacts nontrivially with microwave background
photons.

	The magnitude $\delta T$ of these discontinuities is given by:
\be {{\delta T\over {T}} = {\pm 4\pi G\mu
v_{s}\gamma_{s} {\hat k} \cdot ({\hat v}_s
\times {\hat s})}}, \ee
where ${\vec v}_s$ is the velocity of the string, $\hat k$ is the unit
wave vector of the perturbed photon, and $\hat s$ is a unit vector
along the string.  Due to the effects of compensation \cite{ttb86,m92}
and finite string radius of curvature, the scale over which these
discontinuities persist is about equal to the radius of curvature of
the long string, which according to the scaling solution is
approximated by the causal horizon scale at the time when the photon
interacted with the string.  An observer scanning the sky along a line
which intersects a long string will detect a temperature perturbation
pattern which cuts off at horizon scales due to the above effects; the
degree of sharpness of the cut-off is not well-determined, but
Magueijo has shown that compensation is affected by a gravitational
shock front, which implies a sharp cutoff of the gravitational effects
of long strings at horizon scales \cite{m92b}.  We therefore
approximate the temperature perturbation of a long string as a step
function whose cut-off at the horizon scale is discontinuously sharp,
and whose width is proportional to the horizon scale at the time when
the string was present.  It follows that the predicted temperature
pattern along a line on the sky may be found by superposing the step
functions corresponding to all strings present from $t_{rec}$ to the
present, $t_0$.  This model for the Kaiser-Stebbins effect has the
most discontinuous possible cut-off, so if the actual form of
the string seed function is less discontinuous, our results are too
optimistic.

	The construction of patterns of string-induced cosmic
microwave background fluctuations suggests two interesting classes of
tests for the string model.  The first class involves the properties
of the power spectrum $P(k)$ of the pattern of fluctuations $F(\theta)
= {{{\delta T}\over {T}} (\theta)}$, a function of the angle $\theta$
along an arc of a great circle in the sky.  The Fourier transform of
the pattern is given by \be f(k) = {
\int_{-\infty}^{\infty}{F(\theta) e^{-ik\theta} d\theta} }, \ee and
the power spectrum $P(k)$ is then defined as the ensemble average of
the squared magnitude of the Fourier transform, \be P(k)=<|f(k)|^2>,
\ee where $<>$ denotes ensemble average.

  	The temperature fluctuations detected in the COBE experiment
have been shown to have slightly different spectral indices $n$ for
different methods of calculating $n$, but these values are all
consistent with a scale invariant spectrum.  Bennett {\it et. al.}
find $n\approx {1.4 \pm 0.5}$ \cite{bcobe}, and Wright {\it et. al.}
find $n\approx {1.2 \pm 0.4}$ using a basis other than the spherical
harmonics in order to account for the omission of data in the region
of the galactic plane \cite{w94}.  Scale invariant perturbations are
democratic perturbations for which no particular scale is favored over
others, and whose amplitude is the same for both large and small
scales.  For scale invariant perturbations along an arc on the sky, it
may be shown that \cite{lpsp} $kP(k)=constant$.  The power spectrum is
considered scale invariant when its correlation function, the Fourier
transform of the power spectrum, is constant for all $\theta$.  The
correlation function smoothed on an angular scale $\theta_0$ is given
by
\be
{C(\theta)_{\theta_0}={{<{{\delta T}\over{T}}(\theta_1) {{\delta
T}\over {T}}(\theta_{1} +\theta)>}_{\theta_1}}\approx
{{1 \over{2\pi}} \int{e^{ik\theta}P(k) W(k-k_0) dk}}},
\ee
where $k_0\equiv {\pi \over \theta_0}$
and $W(k-k_0)$ is a filter function filtering out smoothed modes.
For scale invariant perturbations we have
\be
C(0)_{\theta_0} \approx \int P(k) W(k-k_0) dk \approx constant,
\ee
where ${< >}_{\theta_1}$ indicates an average over all ${\theta_1}$
\cite{ge}.  Therefore the $rms$ temperature fluctuations corresponding to
a scale invariant spectrum are independent of the smoothing scale.

	Since string-induced temperature perturbations are produced by
a superposition of step functions whose amplitude is constant on the
average and whose scales vary continuously from a minimum scale equal
to the horizon at $t_{rec}$ to a maximum scale equal to the horizon at
$t_0$, we expect this model to produce a spectrum that is scale
invariant for a wide range of scales.  In the next section we will
test this prediction using Monte Carlo simulations.

	The second class of tests involves the identification of the
non-gaussian features of string perturbations.

	One of the main predictions of models based on inflation is
that the probability distribution of primordial perturbations is
described by a gaussian.  Since the microwave background is a direct
window to these primordial perturbations, inflation predicts that the
probability $Q(\delta)$ for detecting a temperature fluctuation
$\delta={{\delta T}\over T}$ in a given pixel along an arc of the sky
in an experiment is given by \be
Q(\delta)={{1\over{\sqrt{2\pi{\sigma}^2}}}
{exp{(-{{\delta^2}\over{2 \sigma^2}})}}}, \ee where $\sigma={({{\delta
T}\over T})_{rms}}$ is the root mean square temperature fluctuation
obtained from all pixels.

	On the other hand, cosmic string perturbations are produced by
 superposition of seed functions.  Such a superposition in general
leads to a non-gaussian probability distribution for the resulting
perturbations.  However, in the limit of a very large number of
superposed seeds per pixel, the central limit theorem \cite{statref}
predicts that the resulting perturbations will be gaussian to a good
approximation.  The number of superimposed seeds per pixel increases
as the resolution of an experiment decreases.  Thus for large angular
scale experiments such as COBE, whose resolution is $10^\circ$, cosmic
strings predict gaussian behavior of perturbations.  The degree of
non-gaussianity of string perturbations depends crucially on the
parameters of the scaling solution.  For a large number of strings per
horizon volume, $M$, we expect the perturbations to remain gaussian
down to small angular scales.  Numerical simulations of string network
evolution indicate $M \approx 10 $, which may be large enough to wipe
out non-gaussian signatures on scales of one degree or even smaller.

	In this work we use Monte-Carlo simulations to construct
one-dimensional patterns of string-induced temperature fluctuations.
We then use these patterns to test the scale invariance of the
perturbations as well as quantify the dependence of the non-gaussian
features on the resolution, the instrument noise, and the scaling
solution parameters.

	We describe in the next section the approximation used in the
Monte Carlo simulations to construct the temperature fluctuation
patterns for both the string and inflation theories.  We approximate
the fluctuations using step function superposition.  The power
spectrum of the resulting pattern is used to create a random phase
realization pattern which is constructed by assigning random phases to
the Fourier mode amplitudes obtained from the spectrum.  Thus we have
a gaussianized distribution for comparison with the initial one which
has exactly the same power spectrum.  Any differences between the two
patterns are due not to spectrum differences but to the non-gaussian
features of the stringy pattern.  We use Fast Fourier Transform
methods to obtain the power spectra of fluctuation patterns and thus
show explicitly the range of scales over which the scale invariance
persists.  We justify our results via qualitative arguments.

	In Section $3$ we use the constructed temperature patterns to
compare the probability distributions of fluctuations generated by
cosmic strings and by inflation.  We show explicitly that the
non-gaussian features of the string patterns are only evident for
microwave background experiments of very high resolution and find the
minimum resolution required to distinguish the two models.  We also
show that this result is extremely sensitive to instrumental noise;
even a modest amount of gaussian noise is enough to destroy all
non-gaussian features on small scales.

	Finally, in Section $4$ we conclude and discuss interesting
extensions of this project.

\section{Simulations and Spectra }
	We construct patterns of cosmic microwave background
fluctuations by dividing the time between recombination and the
present into a set of discrete expansion time steps and applying
string perturbations at each step.  This type of multiple impulse
approximation for strings was devised by Vachaspati \cite{tv} and
first applied to microwave background fluctuations by Perivolaropoulos
\cite{lpmodel}.  The earliest perturbation and first step occurs at
the recombination time $t_{rec}$, when the angular size $\theta (t_{rec})$
of the causal horizon scale is about $2^{\circ}$.
The angular size of the horizon grows with time according
to \be {\theta(t)\approx {({a(t) \over {a(t_0)}})^{1\over
2}}\approx{t^{1\over 3}}}, \ee
where $a(t)$ is the scale factor of the
universe, and during the matter-dominated era when the string
perturbations would have taken place, $a(t)\approx t^{2\over 3}$.
Each subsequent step occurs when the causal horizon scale $t$ doubles in size.
Namely,
\be
t_{i+1}={2{t_i}} \Longrightarrow {\theta_{i+1} = {2^{1\over 3}
\theta_i}},
\ee
where \be \theta_{1}={\theta({t_{rec}})}\simeq{2^{\circ}}. \ee

	Consider a lattice of $N$ pixels along an arc of the sky with
size $\Theta$.  At the $i$th time step, \be {t_i={2^{i} t_{rec}},
\hspace{1cm} ({t_{rec}} <{t_i} <{t_0})}, \ee and our lattice includes
$q_i$ angular horizon scales, where \be {q_i={\Theta^{\circ} \over
{2^{i/3}\* 2^{\circ}}}}.\ee Let $M \simeq 10$ be the number of long
strings per horizon in the scaling solution as predicted by numerical
simulations of string networks \cite{simul}.  Then the total
number of randomly located step function perturbations $n_i$ to be
superposed on the pixel array during the $i$th expansion step is the
product of the number of horizons and the number of strings per
horizon, \be {{n_{i}}={Mq_i}={M\Theta^{\circ}
\over{2^{1\over3}\* 2^{\circ}}}}.  \ee The angular size $\xi_i$ of
each step function is approximately equal to the horizon size at the
time step when the seed perturbs the lattice, so we have \be
{{\xi_i}={2^{(i-1)\over 3}\* 2^{\circ}}}.  \ee
Since we are not
interested in the overall normalization of the fluctuation patterns,
we choose the amplitude of each step function to be a random number in
the range $[-1,1]$.  This accounts for the random velocities and
orientations of long strings (see Eq. (2)).  By performing the
superposition of seeds as determined by the above quantities for each
time step, we produce a pattern of seed fluctuations.  When the
angular size of the step functions and thus the horizon size becomes
larger than the size of our lattice, the time steps end because such
large seeds affect all of the pixels by the same amount and do not
change ${\delta T}\over T$, the fluctuation pattern.

	After the pattern is constructed, we use FFT methods to find
the Fourier transform of the completed pattern and thus its power
spectrum as given in Eq. (4), by averaging over $100$
realizations.

	For each string fluctuation pattern, we also construct a
corresponding ``random phase realization" (RPR) of the pattern which
serves as an approximation to the type of pattern predicted by
inflation theories.  The RPR is the new temperature fluctuation
pattern obtained by randomizing the phases of the Fourier modes and
then using them to reconstruct a different pattern.
This pattern is the best approximation we can make to inflationary
fluctuations without changing the power spectrum of the pattern.  Thus
the RPR pattern $g(\theta)$ is given by
\be
{g(\theta)={{\sum_{k=-\infty}^{+\infty}}{e^{(ik\theta)}|f(k)|e^{(i{\theta}_k)}}}},
\ee where $\theta_k$ is a random number in the range $[-\pi, \pi]$.
By the central limit theorem, the large number of terms in the sum
(15) implies that the pattern $g(\theta)$ will be random, obeying a
gaussian probability distribution.

	First we demonstrate the scale invariant nature of the
string-induced perturbations in patterns produced by our approximation.
Figure $2$ is a plot of $\log{kP(k)}$ versus $\log{k}$ for an experiment
with resolution $2^{\circ}$ (where $P(k)$ is the ensemble average of
$100$ realizations).  For the slope we found \be {kP(k)\approx k^{-0.16
\pm 0.10}}, \ee which is consistent with a scale invariant spectrum
with $kP(k)$ approximately constant, and spectral index $n\approx {0.84
\pm 0.10}$.  As discussed in the introduction, we expected this result
since the angular scale of the horizon at the recombination time was
only $2^{\circ}$ and the sizes of seeds spanned all larger scales;
thus we see a scale invariant spectrum over a large range of scales.
There are no seeds smaller than the minimum horizon size of
$2^{\circ}$ because any perturbations on photons produced before
$t_{rec}$ are erased by multiple scattering on the plasma.  Thus scale
invariance breaks down for scales less than $2^{\circ}$.
Perivolaropoulos gives a simple analytic derivation of the scale
invariance of the string fluctuations and the range of scales in which
it should persist, {\it i.e.} \be {k \in {[-{\pi\over {\theta_{min}}},
{\pi\over{\theta_{min}}}]}}, \ee which depends on the angular scale of
the last scattering horizon, $\theta_{min}\simeq 2^{\circ}$ \cite{lpsp}.

	We have demonstrated the scale invariance of the
string-induced perturbations, a result that is consistent with COBE
data, but which does not distinguish the predictions of the inflation
model from those of the cosmic string model.  This distinction may be
made by examining the non-gaussian features of the fluctuations in the
statistics of their distributions.

\section{Non-Gaussian Features}
	As discussed in the introduction, most models based on
inflation predict that the probability distribution of microwave
background temperature fluctuations is described by the gaussian of
Eq. (7).  The same kind of gaussian distribution, with
different variance, is predicted for any linear combination of
neighboring pixels, such as the temperature difference between two
adjacent pixels.  In what follows, we use this property of patterns of
temperature differences to examine the non-gaussian features predicted
for patterns of string-induced perturbations created by our Monte
Carlo simulations.

	Two mechanisms tend to suppress the non-gaussian character of
a string fluctuation pattern.  The first is the consequence of the
central limit theorem, which states that a gaussian probability
distribution results from the superposition of a large number of
random variables which may have non-gaussian probability
distributions.  Thus the gaussian is an attractor for all other
probability distributions.  The distribution of string-induced
fluctuations will therefore appear gaussian if the number of seeds
perturbing each pixel is large; this occurs in low
resolution experiments in which the angular size of each pixel is
large, and in theories which predict a large number of seeds per
horizon, {\it i.e.} $M=10$ for strings.

	The second mechanism is instrumental noise, which hides the
non-gaussian signatures by superposing a gaussian distribution of
noise upon them.  The typical signal to noise ratio for microwave
background experiments is about $2$.

	A useful way to determine whether a distribution in a random
variable $x$ obeys a gaussian probability distribution is to compare
the moments of $x$ with the corresponding moments of a gaussian.  The
$n$th moment $\lambda_n$ of a distribution in $x$ is defined as \be
{{\lambda_n}={<x^n>\over{<x^2>^{n\over 2}}} ={ {\int{{x^n}f(x)dx}}
\over{({\int{ {x^2}f(x) dx }})^{n\over 2}}}}, \ee where $f(x)$ is the
probability distribution of the random variable $x$.  The third and fourth
moments are known as the skewness and kurtosis.  For a gaussian
distribution, the skewness is $0$ and the kurtosis is $3$.  In
practice, the skewness and kurtosis are obtained by using $N$ measured
values of the temperature fluctuation, where $N$ would for example be
the number of pixels in an experiment.  Thus, even for a gaussian
distribution the values $\lambda_3 =0$ and $\lambda_4 =3$ would not be
obtained for finite $N$.  Instead, the moments of a discrete
distribution with finite but large $N$ are random variables with
gaussian probability distributions centered about the ideal values of
the moments ({\it i.e.} those for which N approaches infinity), and
with standard deviation $\sigma$ which is proportional to $({1\over
N})^{1\over 2}$ ($\sigma_{\lambda_3} = \sqrt{15\over N},
\sigma_{\lambda_4} = \sqrt{96\over N}$).  Thus a temperature fluctuation
pattern may be characterized as non-gaussian if the measured value of
its kurtosis or skewness lies at least $1\sigma$ from the gaussian
value.

	The moments as defined in terms of the generating function of
a distribution may be used to illustrate the effects of the central
limit theorem on seed-induced fluctuation patterns.  In general the
generating function $M_{\delta}(t)$ is the Laplace transform of the
probability distribution $P(\delta)$ of the variable $\delta$, defined by
\be M_{\delta}(t) = { \int{ e^{t\delta} P(\delta) d\delta }}
\Longrightarrow {<\delta^n>} = {{d^n\over {dt^n}} M(t)
\mid_{t=0}}, \ee where
$t$ is a dummy variable.  Thus the $n$th derivative of the generating
function of a variable is the $n$th moment of the distribution.  The
probability distribution of Eq. (7) may be rewritten through a change
of variable as \be P(\delta) = { {1\over{
\sqrt{2\pi}} } e^{-{{\delta^2}\over 2}} }. \ee Then the generating
function of the gaussian is given by \be M_{\delta}(t) = {
e^{{t^2}\over 2} }, \ee and the definition (19) gives for the skewness
$\lambda_3 = 0$ and for the kurtosis $\lambda_4 = 3$.  An important
property of generating functions is that for two variables $\delta_1$
and $\delta_2$, their generating functions satisfy \be M_{\delta_1 +
\delta_2} = {M_{\delta_1} M_{\delta_2} }.
\ee

	The central limit theorem states that if \be
\delta_n = x_1 + . . . + x_n , \ee
and \be P(x_1) = . . . = P(x_n), \ee
then as $n \longrightarrow \infty$,
\be P(\delta_n) \longrightarrow { {1\over{\sqrt{2\pi\sigma_n^2}}}
 e^{-{{{\delta_n}^2}\over {2\sigma_n^2}}} }. \ee

The temperature fluctuation pattern for gaussian perturbations may be
expanded in Fourier modes as
\be \delta = { \delta T \over T} (\theta) = { \sum_{k= -\infty}^{\infty}
|\delta_k| e^{i \theta_k} e^{i k \theta} }. \ee The gaussian probability
distribution $P(\delta)$ is given in Eq. (20).  For string-induced
perturbations, the temperature fluctuation pattern is given by
\be \delta = {\delta T \over T}(\theta) \approx {\sum_{i=1}^n f_1^{\psi}
 (\theta - \theta_i)},
\ee
where $f_1$ is a seed function (step-like) superposed at random angular
locations $\theta_n$.  The generating function for the
moments of the fluctuation due to the first seed $x_1$ is
\be M_{x_1}(t) = {< e^{t x_1} >} = {(p e^t + p e^{-t} + (1 -2p)e^0)}, \ee
where $p = {\psi \over L}$ is the probability that the step function
perturbation affects a given pixel.  By the property of Eq. (22),
we have for the generating function of moments of the combined
fluctuations of many seeds
\be M_{\delta_n = x_1 +...+x_n} = M_{x_1}*...*M_{x_n} = {(2p \cosh(t) +
(1-2p))}^n. \ee
In the limit $t<<1$, we can expand Eq. (29) to obtain
\be ({1 + {{(2pn) t^2}\over {2n}}})^n,\ee
which for large number of seeds $n>>1$ approaches a gaussian limit,
\be M_{\delta_n} \longrightarrow {e^{{\sigma_{n}^{2} t^2}\over 2}}.\ee
{}From Eq. (28) function we can calculate the moments of the seed
fluctuation pattern according to Eq. (19):
\be \lambda_2 = \sigma_n^2 = {d^2 \over {dt^2}} M_{\delta}(t) = 2np, \ee
\be \lambda_3 = {d^3 \over {dt^3}} M_{\delta}(t) = 0, \ee
\be \lambda_4 = {{d^4 \over {dt^4}} M_{\delta}({t \over \sigma_n})} = {3 +
{{1-6p}\over{2np}}}
 \longrightarrow 3 .\ee Thus we have shown that in accordance with the
central limit theorem, the moments of the string-induced fluctuation
pattern approach the values of the corresponding moments of the
gaussian probability distribution for large numbers of seeds.

	We use the moments to compare our string-induced fluctuation
patterns with their RPRs and thus find the values of parameters such
as signal to noise ratio and experimental resolution which allow us to
detect the non-gaussian nature of a string pattern.  As described in
Section $2$, we create the patterns of string fluctuations from
superposition of step function perturbations on a one-dimensional
array of $512$ pixels; the inflationary patterns follow by randomizing
the phases of the Fourier modes of the string fluctuations (the RPR),
but retaining the same power spectrum as that of the string pattern.
We varied the values of angular resolution (or pixel size) and signal
to noise ratio, and constructed $100$ patterns for each case.  The
$100$ patterns were used to calculate the means and standard
deviations of $\lambda_3$ and $\lambda_4$ for each experiment.

	Instrumental noise was introduced by adding to the temperature
fluctuation value of each pixel an uncorrelated gaussian random
component such that the ratio of the standard deviation of this
component to the standard deviation of the true signal (temperature
fluctuation) was set equal to the inverse of the signal to noise ratio
(${(s/n)}^{-1}$).  (Where the noise is zero we do not define $s/n$.)
Although measurements of microwave background radiation are plagued by
correlated non-gaussian noise from our galaxy as well as instrumental
noise, the effects of galactic noise are usually minimal compared to
the latter because galactic noise can be modeled and subtracted from
the data.  In addition, even full sky maps such as those of COBE
exclude data from the galactic plane.  Residual galactic noise would
have a small non-gaussian influence on the kurtosis of the
fluctuations, but this effect would be the same at all experimental
resolutions because galactic noise is continuous.  The galactic
non-gaussian component in the kurtosis of the temperature fluctuations
thus remains constant with improving resolution, in contrast to the
kurtosis of string-induced fluctuations which rapidly increases with
increasing resolution because of the discontinuous nature of string
seeds, as we will show.  Thus the non-gaussian character of galactic
noise would be easily distinguishable from that of string-induced
fluctuations at high resolutions.  Since adding galactic noise would
only shift the kurtosis of a fluctuation pattern up by the same small
amount at all resolutions, we neglect it, and include only
uncorrelated gaussian noise in our model.

	The step functions superposed to create the string
fluctuations are antisymmetric, so it is equally likely for a
temperature perturbation to be positive or negative.  Thus the
skewness predicted for the string pattern is zero, since this moment
measures the symmetry of a distribution.  Since zero is also the ideal
value of the gaussian skewness, the skewness is not useful for
distinguishing the string and gaussian patterns.  We therefore focus
on the statistics of the kurtosis $\lambda_4$ in what follows.  (It has
been shown that for higher moments of string patterns, the
non-gaussian character is more easily detected than for the kurtosis,
but the errors for higher moments increase faster than the degree of
non-gaussianity, so we do not consider them \cite{lpmom}.)

	Figure $3$ is a plot of kurtosis versus angular resolution
with zero noise for both string and gaussian temperature fluctuation
patterns.  The values of $\lambda_4$ for both the seed and gaussian
patterns are nearly the same, approximately the gaussian value of $3$,
for even the highest resolutions considered (0.5 arcmin).  Thus even
under the best experimental conditions (very high resolution and zero
noise), this kurtosis test cannot distinguish between the two
temperature fluctuation patterns.  The number of seeds falling on the
array is large enough to obscure the non-gaussian character of the
seed perturbations as the central limit theorem predicts.

	In order to improve our chances of observing the non-gaussian
signatures of string perturbations, we cannot lower the number of
seeds perturbing each horizon ($M=10$) at each expansion step, since
it is fixed by numerical simulations of the string scaling solution.
We can, however, decrease the influence of the other factor causing
large numbers of seeds to perturb a given pixel, the width of each
seed step function; each step function affects all of the pixels
within its horizon, and by the end of the simulation one perturbation
spans the whole pixel array.  The way to eliminate this effect is to
examine the difference between temperature fluctuations in adjacent
pixels, which is proportional to the derivative of the pattern.  In
this case the seeds superposed are not step functions but delta
functions localized at the discontinuities induced by the step
functions.  We thus dramatically decrease the number of perturbations
which influence each pixel, improving our chance of observing any
non-gaussian characteristics in the string patterns.

	The success of this method in revealing the difference between
the seed and gaussian fluctuations is demonstrated in Figure $4$, in
which we plot the kurtosis of the gradient of fluctuations versus
resolution, and compare the patterns and their RPRs with zero noise.
The plot shows that the non-gaussianity of the seed fluctuations is
clear only for very good resolutions of less than $0.1^{\circ}$ ($6$
arcminutes).  The error bars indicate $1\sigma$ deviations; since the
seed and gaussian errors do not overlap at high resolutions, we can
distinguish fairly well between the two patterns for experiments with
such resolution.

	However, any realistic experiment will necessarily include
some level of instrumental noise.  Will the predicted non-gaussian
features persist when the effects of noise are taken into account?

	We addressed this question by adding a modest amount of noise
($s/n=3.3$) to the temperature fluctuation maps and their RPRs and
then constructing the temperature difference maps.  In Figure $5$, a
plot of $\lambda_4$ versus angular resolution, we show that the noise
completely removed the non-gaussian features for all resolutions.
Though examining the temperature differences reduces the effect of too
many seeds leading to gaussianity in the moments, it also greatly
reduces the amount of ``signal" in the average pixel, since it only
counts the delta function spikes of the seed.  The amplitude of the
noise added depends on the standard deviation of the temperature
fluctuation pattern from its mean, however, not the deviation of the
temperature gradient pattern.  It thus overwhelms the small amount of
non-gaussian information in the gradient pattern and produces a
pattern composed mainly of gaussian noise whose moments are highly
gaussian.

	We conclude that unless the amount of instrumental noise is
unrealistically low ($s/n\simeq 10$ or greater), it is extremely
difficult to detect the non-gaussian character of the cosmic string
fluctuations using tests based on the skewness or kurtosis.

	An obvious way to amplify the non-gaussian features predicted
by models based on topological defects is to decrease the number $M$
of superposed seeds per horizon scale.  To do this we must consider
models based on topological defects other than strings, since for
strings $M$ is fixed by simulations to be about $10$.  In the case of
textures, for instance, we expect to find that tests based on the
kurtosis will be more successful than they are with strings, because
simulations predict that their value of $M$ is about $0.05$.  While
$10$ strings appear in each horizon, only $1$ texture forms and
collapses in every $20$ horizons at a given time.  For non-string
types of seeds, the structures of the perturbations they cause are not
the same as the step function model appropriate for strings; the forms
of the other seed perturbations are in general not discontinuous, so
their probability distributions are more gaussian than that of the
superposed step functions.

	However, keeping in mind that we may be overestimating the
non-gaussian features, we may use the step functions to obtain a rough
answer to the following question: What is the maximum number of seeds
$M$ per horizon scale for a model in order for this model to have
detectable non-gaussian features for resolutions of about $1$
arcminute and $s/n \approx 5$?

	In Figure $6$, a plot of kurtosis $\lambda_4$ versus $M$
for experiments with $s/n=5$ and resolution of $1$ arcminute shows
that the maximum $M$ for which the kurtosis is clearly not gaussian is
$0.1$.  This suggests that it may be possible to detect non-gaussian
signatures of the texture model for experiments with resolution
$\theta_{min}\leq 1$ arcminute and $s/n \geq 5$.

\section{Conclusion and Outlook}
	We have compared the temperature fluctuation and temperature
gradient patterns created by Monte Carlo simulations based on a simple
model for cosmic string fluctuations in the cosmic microwave
background.  We confirmed that our patterns have scale invariant power
spectra over a wide range of scales, in agreement with COBE
observations and the predictions of the topological defect and
inflation theories; we showed that the kurtosis test can detect the
non-gaussian features of the seed patterns, but for strings this test
fails when we include even a small amount of instrumental noise
because of the large number of seeds per horizon $M=10$.  For seed
perturbations of the form of the string step function, the maximum
value of $M$ for which we might expect to detect the non-gaussian
characteristics of a pattern is about $0.1$, but this is an optimistic
figure since other topological defects generally induce seed
perturbations which are smoother than those of strings.

	Since the moments are capable of distinguishing seed and
gaussian patterns only in experiments with unrealistically low amounts
of noise, they are not a viable test with which to analyze observed
temperature fluctuation patterns and determine the nature of the
primordial perturbations in the CMB.  The study of other tests on the
statistics of seed and gaussian patterns, such as the correlation of
the Fourier phases of the power spectra, may produce more positive
results.

\newpage
\section{Figure Captions}
\indent

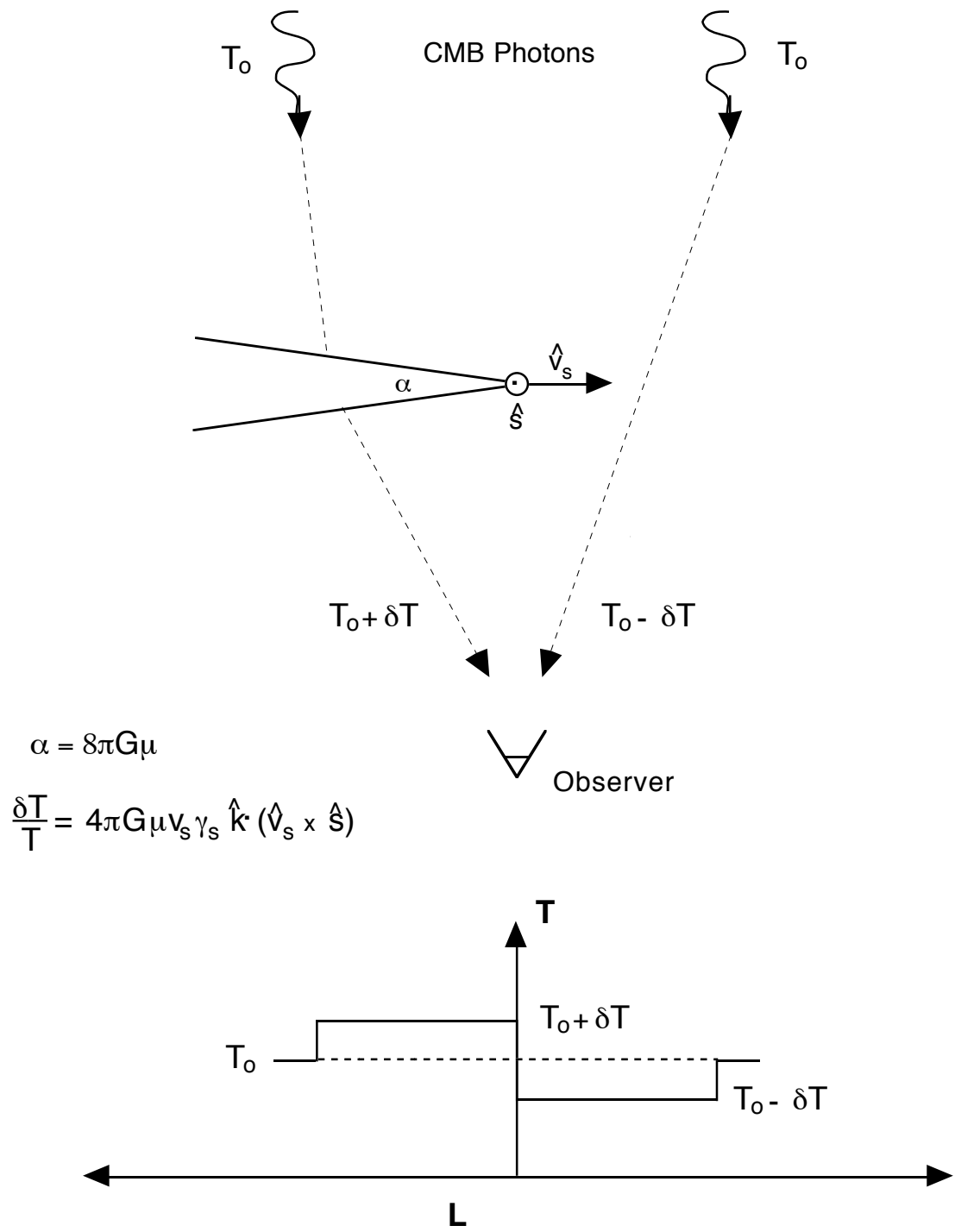
{\bf Figure 1:} Kaiser-Stebbins Effect.  A cosmic string with a
velocity component perpendicular to the observer's line of sight
induces a temperature perturbation ${\delta T} \over T$ in microwave
background photons crossing its path within a horizon scale.

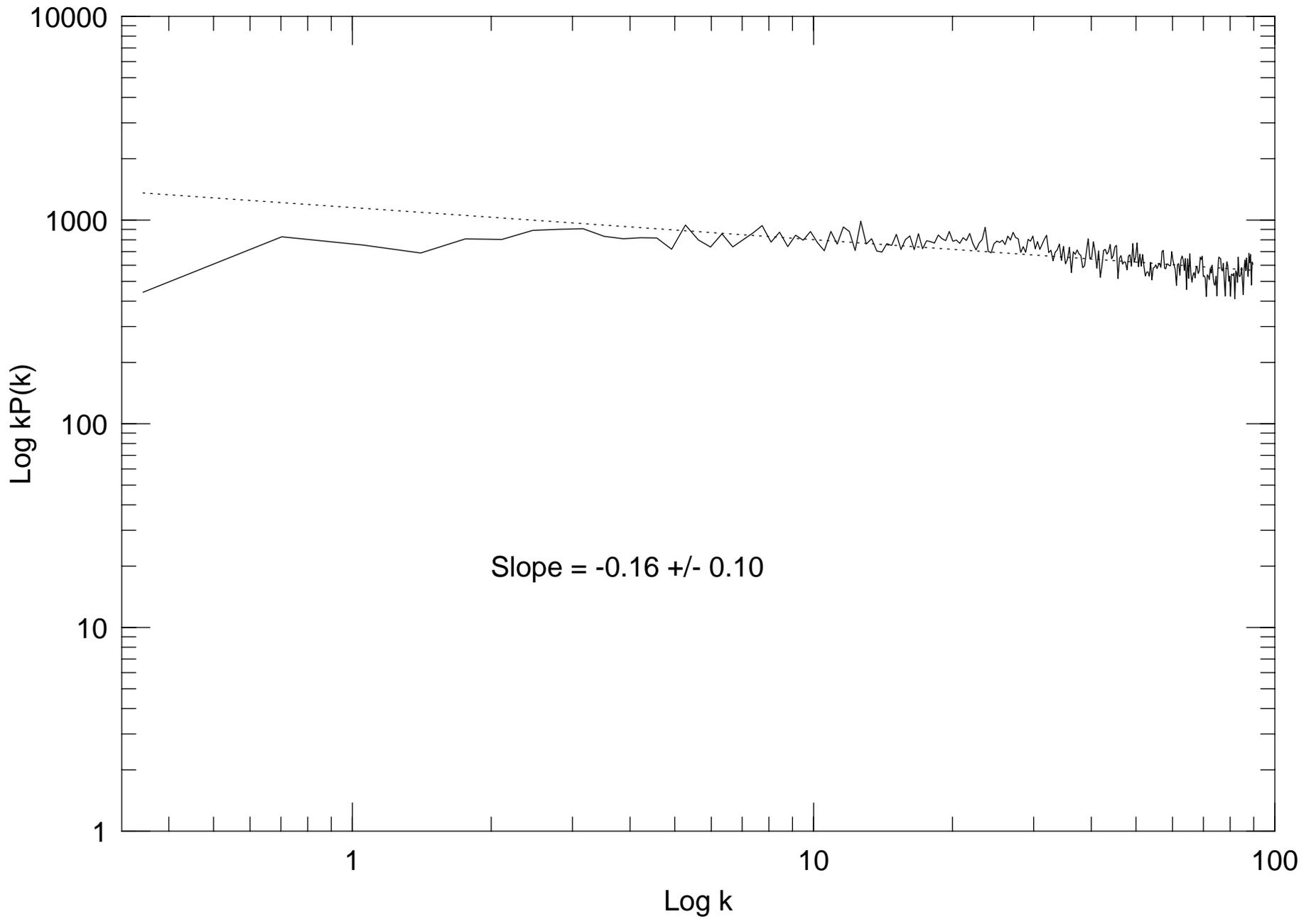
{\bf Figure 2:} This plot of $\log {kP(k)}$ versus $\log {k}$ shows
the large range over which the fluctuation pattern is scale invariant.
It has a nearly constant slope of $-0.16 \pm 0.10$, which gives the
spectral index $n \approx {0.84 \pm 0.10}$.

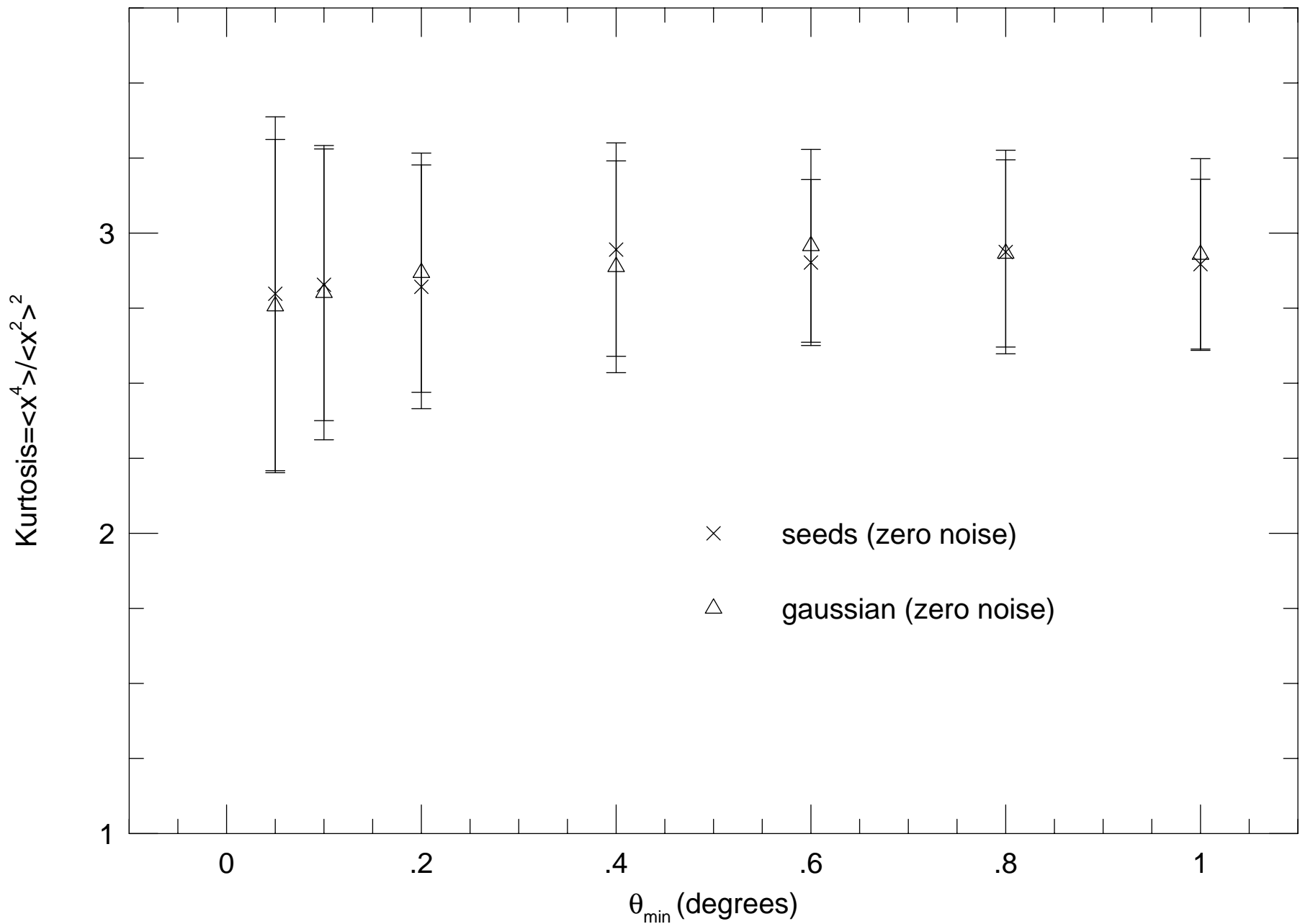
{\bf Figure 3:} Plot of kurtosis of temperature fluctuations versus
angular resolution ($\theta_{min}$) with zero noise.  Even at the
highest resolutions, this test cannot distinguish the gaussian from the
string pattern.

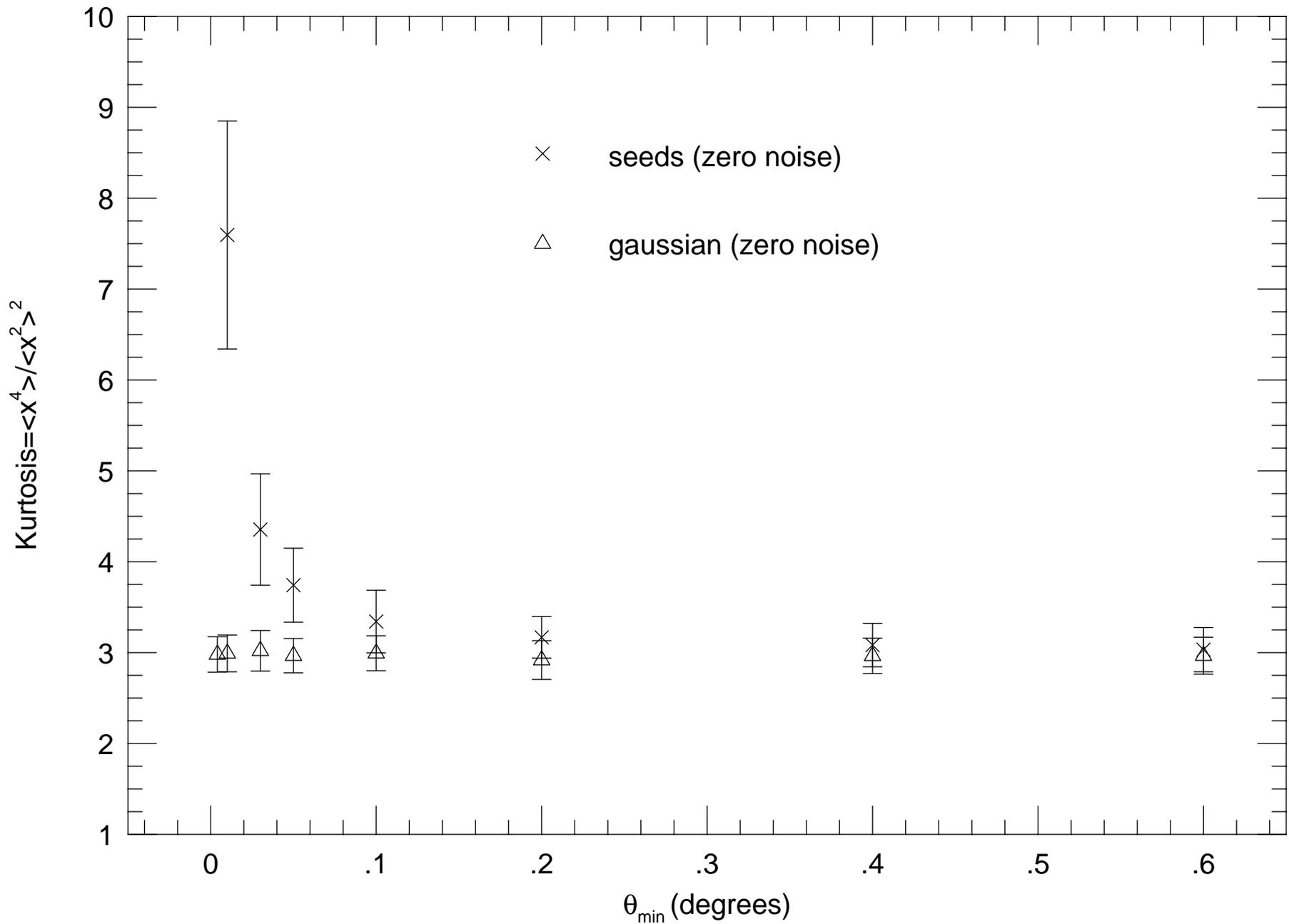
{\bf Figure 4:} Plot of kurtosis of temperature gradient of
fluctuations versus angular resolution.  For high resolutions and
zero noise, this test can distinguish between seed and gaussian
patterns.

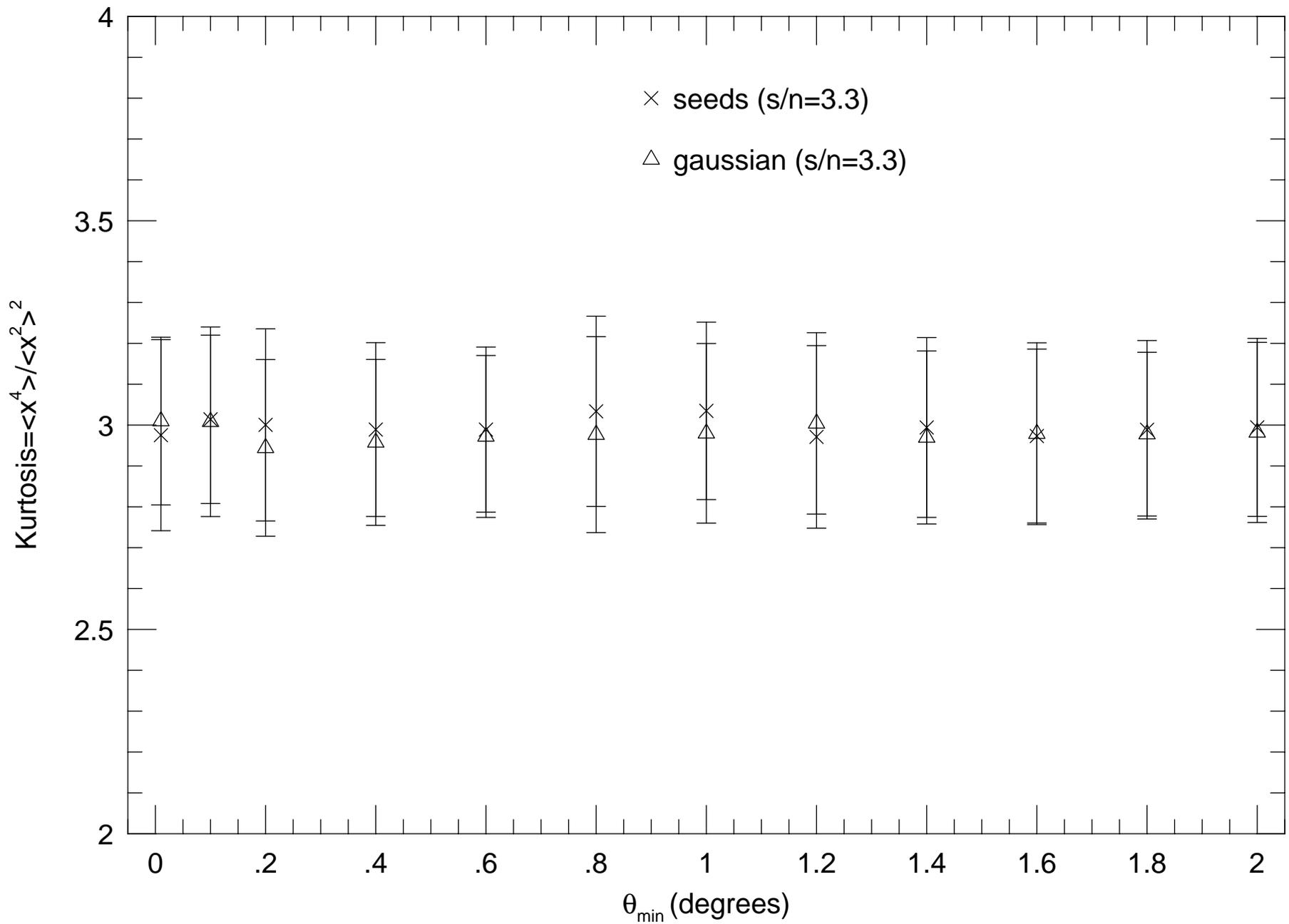
{\bf Figure 5:} Same as Figure $4$ but with gaussian instrumental noise
added.  A signal to noise ratio of $3.3$ destroys all trace of the
non-gaussian features of the seed pattern.

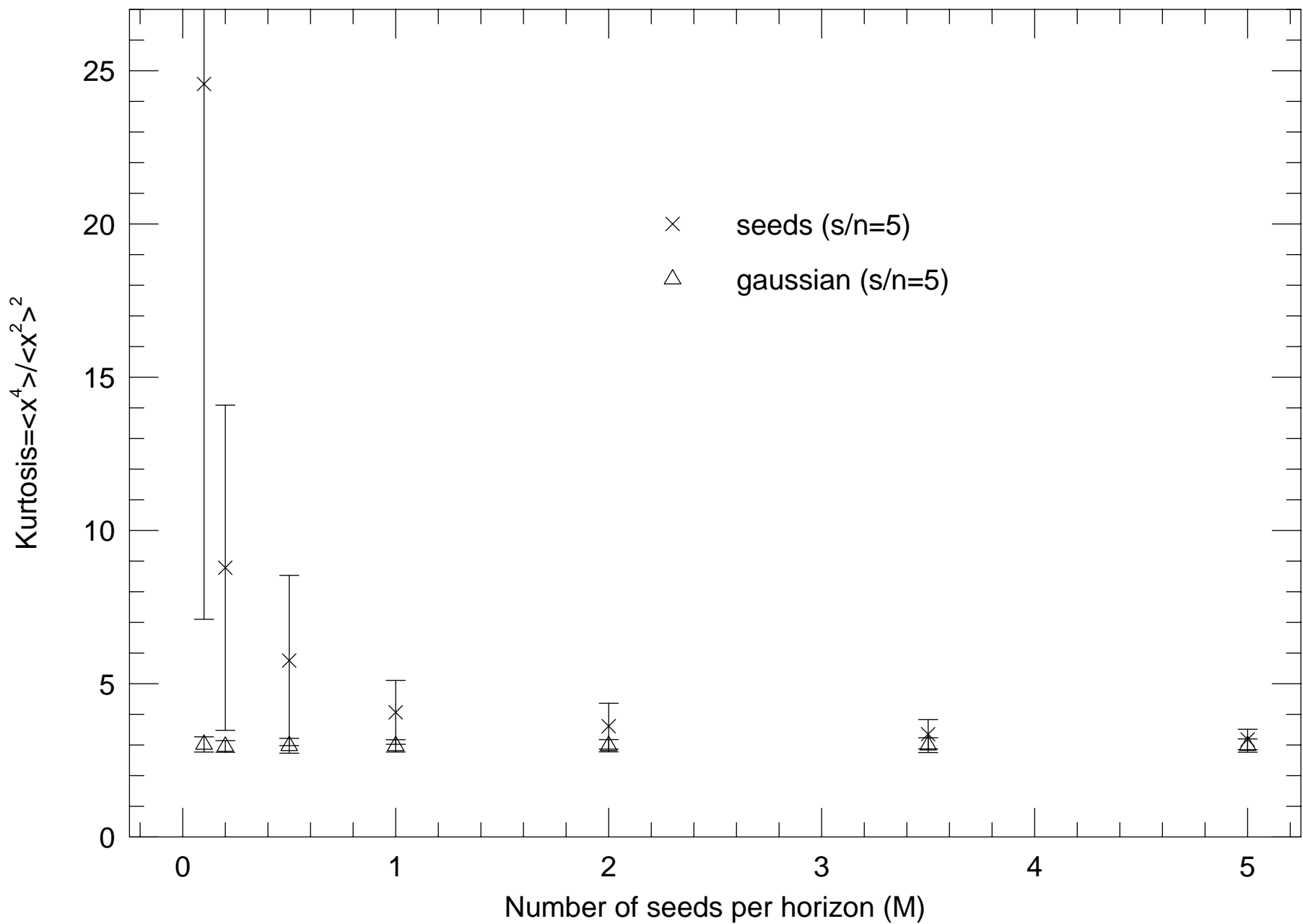
{\bf Figure 6:} Plot of kurtosis versus number of seeds per horizon
for seeds of the form of step functions.  For the smallest values of $M$
the kurtosis of string patterns deviates most from the gaussian,
suggesting that fluctuation patterns for seeds with smaller $M$ values
such as textures would be more easily distinguished from the gaussian
than string patterns.

\end {document}